\documentclass[preprint,5p]{elsarticle}

\usepackage{graphicx}% Include figure files
\usepackage{dcolumn}% Align table columns on decimal point
\usepackage{bm}% bold math
\usepackage[colorlinks,citecolor=blue,hyperindex,breaklinks]{hyperref}
\usepackage{breakurl}
\usepackage[colorinlistoftodos]{todonotes}
\usepackage{ulem}
\usepackage{amsmath,amssymb}

\usepackage{wasysym}

\usepackage{verbatim}

\usepackage{orcidlink}
\usepackage{textcomp}

\journal{Physics Letters B}

\newcommand{\rmd}{\mbox{d}}

\begin{document}
    
    \begin{frontmatter}
        
        \title{Moments of inertia of rare-earth nuclei and the nuclear time-odd mean fields within exact solutions of the adiabatic theory}
        
        \author[UoY]{Xuwei Sun\orcidlink{0000-0002-0130-6269}}
        \ead{xuwei.sun@york.ac.uk}
        
        \author[UoY,UoW]{Jacek Dobaczewski\orcidlink{0000-0002-4158-3770}}
        
        \author[UoJ]{Markus Kortelainen\orcidlink{0000-0001-6244-764X}}
        
        \author[VACC,HBNI]{Jhilam Sadhukhan\orcidlink{0000-0003-1963-1390}}
        
        \author[UoY]{Adri\'an S\'anchez-Fern\'andez\orcidlink{0000-0003-3502-6668}}
        
        \author[UoY]{Herlik Wibowo\orcidlink{0000-0003-4093-0600}}
        
        \address[UoY]{School of  Physics, Engineering and Technology, University of York,  Heslington, York YO10 5DD, United Kingdom}
        \address[UoW]{Institute of Theoretical Physics, Faculty of Physics, University of Warsaw, ul. Pasteura 5, PL-02-093 Warsaw, Poland}
        \address[UoJ]{Department of Physics, University of Jyväskylä, P.O. Box 35, FI-40014 Jyväskylä, Finland}
        \address[VACC]{Physics Group, Variable Energy Cyclotron Centre, 1/AF Bidhan Nagar, Kolkata-700064, India}
        \address[HBNI]{Homi Bhabha National Institute, Anushakti Nagar, Mumbai-400094, India}

        \begin{abstract}
            We systematically analyse the nuclear moments of inertia determined within the Skyrme and Gogny density functional theories.
            The time-odd mean fields generated by collective rotation are self-consistently determined by a novel exact iterative solution of the adiabatic time-dependent Hartree-Fock-Bogoliubov (ATDHFB) equations. Although details of the results depend on the functional used, the calculated moments of inertia are in good overall agreement with the experimental data, with no adjustable parameters. To show the essential importance of the time-odd mean fields, we compared the ATDHFB moments of inertia with those obtained from the Inglis-Belyaev formula.
            For Skyrme density functionals, we find strong correlations between the effective mass and the impact of the time-odd mean fields on the rotational and vibrational collective inertia.
        \end{abstract}
        
        \begin{keyword}
            rotational moment of inertia \sep vibrational collective inertia \sep density functional theory
            \sep adiabatic approximation
            \sep effective mass
        \end{keyword}
        
    \end{frontmatter}

    Collective motion provides a deep insight into the nuclear reaction mechanism and the quantum dynamics of many-nucleon systems. 
    The low-energy collective motion is particularly interesting, usually accompanied by large electromagnetic moments and transition rates.
    To fully understand them, we must scrutinise the collective inertia, which measures the quantum system's resistance to collective motion. 

    Studies of collective motion are important in nuclear physics and attract experimental and theoretical attention. 
    The ATDHFB method~\cite{BARANGER1978123}, established within the self-consistent nuclear density functional theory (DFT), is one of the most prominent microscopic approaches to describe the inertia of large-amplitude low-energy collective modes.
    The adiabatic assumption is valid for collective motion when it is much slower than the single-particle motion of individual nucleons.
    It also bridges the microscopic many-body theory and phenomenological collective models solely based on collective variables.

    An exact determination of the ATDHFB inertia requires an inversion of the two-body stability matrix~\cite{Ring2004book}. For deformed and/or superfluid systems, such explicit inversion requires a prohibitive computational cost because of the large matrix dimensions involved. 
    Due to such numerical difficulties, the ATDHFB method was often used in the cranking approximation~\cite{Baran2011}, which neglects the dynamical residual interactions generated by the collective motion.
    Collective inertia obtained in the cranking approximation is equivalent to that calculated using the Inglis~\cite{Inglis1956PRC} or Inglis-Belyaev (IB)~\cite{BELIAEV1961NP} formula.
    The situation is further worsened in the perturbative cranking approximation~\cite{Goutte2005, Baran2011}, where the inertia is calculated via energy-weighted moment tensors that involve only the diagonal element of the stability matrix.
    Such an omission of the residual interaction, especially in the time-odd mean-field channels, leads to a violation of the Galilean symmetry, which has severe consequences, for example, resulting in the incorrect translational mass, see, e.g., Ref.~\cite{Wen2022}.
    
    Despite the inconsistency caused by neglecting the time-odd mean fields, the cranking approximation has been used in theoretical studies of collective inertia, ranging from calculations of low-energy spectra~\cite{Nik2009PRC} to evaluations of nuclear fission half-lives~\cite {Sadhukhan2013}. Such an inconsistency led to the enhancement factor adopted to correct the collective inertia obtained from the cranking approximation~\cite{Libert1999PRC}.
    To alleviate the inconsistency, a method based on the expansion of the inertia matrix was developed to solve the ATDHFB equation in the absence of pairing, which can evaluate the rotational moment of inertia with high precision of the order of 1\%~\cite{ZPLI2012PRC}.
    Recently, the similarity of the ATDHFB method and the finite-amplitude method (FAM) at zero frequency was realised, and a quasiparticle FAM on top of the Skyrme energy density functional was developed to evaluate the Thouless-Valatin rotational moment of inertia~\cite{Petrik2018PRC}. With the pairing strength adjusted to the experimental pairing gaps, the rotational moment of inertia in several axially deformed nuclei from the rare-earth and heavy-actinide regions was well reproduced, and the large and mass-dependent deviations from the values evaluated by the cranking approximation were confirmed. The FAM was also used in Ref.~\cite{(Was24)} for the full five-dimensional quadrupole adiabatic inertia.
    
    This letter presents a novel iterative method for solving the ATDHFB equations exactly and efficiently. We implemented this method within the non-relativistic nuclear DFT framework and determined the adiabatic time-odd mean fields self-consistently. Below, we present a systematic analysis of the nuclear moments of inertia and a preliminary study of vibrational inertia in heavy deformed nuclei.
    
    The iterative ATDHFB method is outlined as follows.\footnote{We presented details of the iterative ATDHF method (without pairing) in our recent conference publication~\cite{(Sun25a)}.}
    In the nuclear DFT with pairing, the many-body wavefunction of the nuclear system is mapped onto a one-body quasiparticle density $\mathcal{R}$, whose time-evolution obeys the time-dependent Hartree-Fock-Bogoliubov equation~\cite{Ring2004book, Nakatsukasa2016RMP},
    \begin{equation}\label{TDHFB}
        i\hbar\dot{\mathcal{R}}(t) = \left[\mathcal{H}(t),\mathcal{R}(t)\right],
    \end{equation}
    where the dot represents the derivative over time, $\dot{\mathcal{R}}(t)={\rmd\mathcal{R}(t)}/{\rmd{}t}$. We see that the static (time-independent) solution $\mathcal{R}_0$ then implies $\left[\mathcal{H}_0,\mathcal{R}_0\right]=0$.
    
    At every time $t$, the one-body quasiparticle Hamiltonian $\mathcal{H}(t)=\partial{}E[\mathcal{R}(t)]/\partial\mathcal{R}(t)$ is the derivative of the energy $E[\mathcal{R}(t)]$ of the nuclear system over the generalised density matrix $\mathcal{R}(t)$. In non-relativistic nuclear DFT, it is typically determined using either the zero-range Skyrme functional~\cite{Skyrme1958} or the finite-range Gogny functional~\cite{Robledo2018}.
    
    In the adiabatic limit, we assume that the system progresses in time along the given collective path $q(t)$, namely $E[\mathcal{R}(t)]\equiv{}E[\mathcal{R}(q(t))]$. Here, for simplicity, we consider only one collective coordinate $q$. In addition, we assume that the velocity $\dot{q}(t)$ of the collective motion is so small that the nucleus remains almost in static equilibrium at all times, that is,
    \begin{eqnarray}
       && \mathcal{R}(q(t))\simeq\mathcal{R}_0(q(t))+\dot{q}(t)\tilde{\mathcal{R}}_1(q(t))+\ldots , \\
       && \mathcal{H}(q(t))\simeq\mathcal{H}_0(q(t))+\dot{q}(t)\tilde{\mathcal{H}}_1(q(t))+\ldots , \\
       && \left[\mathcal{H}_0(q(t)),\mathcal{R}_0(q(t))\right] \simeq 0,
    \end{eqnarray}
    where the equalities hold only up to terms linear in collective velocity $\dot{q}(t)$.
    All these densities and mean fields depend on time only parametrically; that is, through the dependence on time of $q(t)$. This allows us to omit, for clarity, the implicit arguments $q(t)$ from the expressions that follow. In particular, the ATDHFB equation, which is the first-order part of the TDHFB equation~(\ref{TDHFB}), takes the following form,
    \begin{equation}\label{ATDHFB_EQ}
        i\hbar  \frac{\partial\mathcal{R}_0}{\partial{q}} = [\mathcal{H}_0,\tilde{\mathcal{R}}_1] + [\tilde{\mathcal{H}}_1,\mathcal{R}_0],
    \end{equation}
    where the derivative over $q$ replaced the time derivative, $\dot{\mathcal{R}_0} = \dot{q} \frac{\partial\mathcal{R}_0}{\partial{q}}$, and thus the ATDHFB equation became independent of the velocity $\dot{q}$.

    Within the adiabatic approximation~\cite{BARANGER1978123}, the quasi-static zero-order densities and mean field are assumed to be time-even, $\mathcal{T}^+\mathcal{R}_0\mathcal{T}=\mathcal{R}_0$ and $\mathcal{T}^+\mathcal{H}_0\mathcal{T}=\mathcal{H}_0$. In contrast, the first-order corrections are assumed to be time-odd, $\mathcal{T}^+\tilde{\mathcal{R}}_1\mathcal{T}=-\tilde{\mathcal{R}}_1$ and $\mathcal{T}^+\tilde{\mathcal{H}}_1\mathcal{T}=-\tilde{\mathcal{H}}_1$. 

    For the collective rotation, the collective path is straightforward. It corresponds to the deformed state rotated in space by angle $\theta$, that is, $|\Phi(\theta)\rangle=\exp\left(i\theta\hat{I}_x\right)|\Phi(0)\rangle$, about the selected axis $x$, and then using $q\equiv\theta$. In this case, the density derivative $\frac{\partial\mathcal{R}_0}{\partial{q}}$ is given by the angular momentum $\hat{I}_x$ matrix elements, which can be calculated analytically~\cite{Sun2025}. For the collective vibration, it can be evaluated by the numerical differentiation of the HFB densities~\cite{(Sun25a)} obtained by constraining, e.g., values of the axial quadrupole moment, $\delta\langle\Phi|\hat{H}-\lambda\hat{Q}_{20}|\Phi\rangle=0$, and then using $q\equiv\langle\hat{Q}_{20}\rangle$. This differs from Ref.~\cite{(Was24)}, where the density derivatives were determined after solving the local quasiparticle random phase approximation equations.

    Since $\tilde{\mathcal{H}}_1$ up to the first order depends linearly on $\tilde{\mathcal{R}}_1$, the ATDHFB equation~(\ref{ATDHFB_EQ}) is a linear equation for $\tilde{\mathcal{R}}_1$ in many dimensions, which necessitates an impractical inversion of an exceedingly large stability matrix~\cite{Ring2004book}. To circumvent this problem, the novel method proposed here transforms the ATDHFB equation~(\ref{ATDHFB_EQ}) into an iterative fixed-point equation,
    \begin{equation}\label{ATDHFB_EQ_it}
        i\hbar\frac{\partial\mathcal{R}_0}{\partial{q}} = [\mathcal{H}_0,\tilde{\mathcal{R}}_1^{(n)}] + [\tilde{\mathcal{H}}_1^{(n-1)},\mathcal{R}_0],
    \end{equation}
    where $\tilde{\mathcal{H}}_1^{(n-1)}$ is evaluated for $\tilde{\mathcal{R}}_1^{(n-1)}$ and $\tilde{\mathcal{R}}_1^{(0)}\equiv\tilde{\mathcal{H}}_1^{(0)}\equiv0$. 
    Then, at each iteration $n=1,2,\ldots{N}$, the collective inertia can be evaluated as
    \begin{equation}\label{adb_moi}
        \mathcal{M}^{(n)}=\frac{i\hbar}{2}\,\textrm{Tr}\left(\frac{\partial\mathcal{R}_0}{\partial{q}}[\mathcal{R}_0,\tilde{\mathcal{R}}_1^{(n)}]\right),
    \end{equation}
    and the iteration stops when $\mathcal{M}^{(n+1)}\simeq\mathcal{M}^{(n)}$ within a prescribed suitable precision.  In practical applications, about two dozen iterations suffice.
    
    A practical solution of the iterative equation~(\ref{ATDHFB_EQ_it}) can be most easily obtained
    in the quasiparticle basis, where it reads, cf.~Refs.~\cite{(Dob81a),Baran2011},
    \begin{equation}\label{ATDHFB_EQ_it_qp}
        Z_{\mu\nu}^{(n+1)} = \frac{1}{E_{\mu}+E_{\nu}}(i\hbar{}F - E_1^{(n)})_{\mu\nu},
    \end{equation}
    where the antisymmetric matrices $Z=\varphi^{+}\tilde{\mathcal{R}}_1\chi$, $F =  \varphi^{+}\frac{\partial \mathcal{R}_0}{\partial q}\chi$, and $E_1 = \varphi^{+}\tilde{\mathcal{H}}_1\chi$ are defined by the standard quasiparticle ($E_\mu>0$) and quasihole ($E_\mu<0$) wave functions $\chi$ and $\varphi$, respectively, having the form
    \begin{equation}\label{ATDHFB_wf}
        \chi =
        \bigg(
        \begin{array}{c}
            A\\
            B\\
        \end{array}
        \bigg),\quad
        \varphi =
        \bigg(
        \begin{array}{c}
            B^*\\
            A^*\\
        \end{array}
        \bigg),
    \end{equation}
    with $A$ and $B$ defining the general Bogoliubov transformation.
    Positive quasiparticle energies ($E_\mu,E_\nu>0$) are taken in Eq.~(\ref{ATDHFB_EQ_it_qp}).
    
    To proceed with the iteration in Eq.~(\ref{ATDHFB_EQ_it_qp}), $E_1^{(n)}$ is computed from $Z^{(n)}$, which closely resembles the standard iteration of the HFB equation.
    Indeed, at each iteration, we first perform the singular value decomposition (SVD) of the antisymmetric matrix $Z=U \Omega V^{+}$, where $U$ and $V$ are unitary and $\Omega$ is diagonal and non-negative; that is, $\Omega_{\mu\nu}=\delta_{\mu\nu}\omega_\mu$ and $\omega_\mu\geq0$. Next, we observe that the eigenequation for the Hermitian matrix $\mathcal{Y}=\left(\begin{array}{cc}0&Z\\Z^+&0\end{array}\right)$ (the first-order density matrix $\tilde{\mathcal{R}}_1$ expressed in the quasiparticle basis)
    takes the following form:
    \begin{equation}\label{ATDHFB_eig}
        \mathcal{Y}\bigg(
        \begin{array}{c}
            U\\
            \pm V
        \end{array}
        \bigg)
        =\bigg(
        \begin{array}{c}
            \pm U \Omega\\
            V \Omega\\
        \end{array}
        \bigg)
        =\pm \bigg(
        \begin{array}{c}
            U\\
            \pm V\\
        \end{array}
        \bigg)\Omega.
    \end{equation}
    This indicates that the eigenvectors of $\mathcal{Y}$ occur in pairs of opposite eigenvalues, and the matrix $\mathcal{Y}$ can be represented as the sum of its eigenvectors\footnote{Note that the eigenvectors in Eq.~(\protect\ref{ATDHFB_eig}) are normalised to 2. This is so because, in the standard definition of the SVD, the columns of the unitary matrices $U$ and $V$ are normalised to 1.} as follows:
    \begin{equation}\label{ATDHFB_sum}
        \mathcal{Y}=\tfrac{1}{2}
        \bigg(\begin{array}{cr}
            U &  U\\
            V & -V\\
        \end{array}\bigg)
        \bigg(\begin{array}{cc}
            \Omega &  0     \\
            0      & -\Omega\\
        \end{array}\bigg)
        \bigg(\begin{array}{cr}
            U^+ &  V^+\\
            U^+ & -V^+\\
        \end{array}\bigg).
    \end{equation}

    Therefore, the first-order correction matrix, $\tilde{\mathcal{R}}_1=\mathcal{A}\mathcal{Y}\mathcal{A}^+$, for $\mathcal{A}=\left(\begin{array}{cc}A&B^*\\B&A^*\end{array}\right)$, can be obtained through the inverse Bogoliubov transformation, and thus it can be represented in its final form as,
    \begin{eqnarray}\label{ATDHFB_final}\hspace*{-1cm}
        \tilde{\mathcal{R}}_1
        &=&
        \bigg(\begin{array}{cr}
            A_1 &  {B_1'}^*\\
            B_1 &  {A_1'}^*\\
        \end{array}\bigg)
        \bigg(\begin{array}{cc}
            \Omega &  0     \\
            0      & -\Omega\\
        \end{array}\bigg)
        \bigg(\begin{array}{cr}
            A_1^+ &  B_1^+\\
            {B_1'}^T &  {A_1'}^T\\
        \end{array}\bigg)
        \nonumber \\
        &=&
        \bigg(\begin{array}{cr}
            \rho_1 &  \kappa_1\\
       -\kappa_1^* & -\rho_1^*\\
        \end{array}\bigg) =
        \chi_1\Omega\chi_1^+
       -\varphi_1\Omega\varphi_1^+ ,
   \end{eqnarray}
    where $\rho_1$ and $\kappa_1$ are the first-order corrections to the standard density matrix and pairing tensor~\cite{Ring2004book}, and
    \begin{eqnarray}\label{ATDHFB_wf1}
    \chi_1 &=&
    \begin{array}{l}
       \left(\begin{array}{c}A_1 \\ B_1 \end{array}\right) = \frac{1}{\sqrt{2}}
       \left(\begin{array}{c}A U + B^* V \\
                             B U + A^* V\end{array}\right),
    \end{array} \\ \label{ATDHFB_wf2}
    \varphi_1 &=&
    \begin{array}{l}
       \left(\begin{array}{c}{B_1'}^* \\ {A_1'}^* \end{array}\right) = \frac{1}{\sqrt{2}}
       \left(\begin{array}{c}A U - B^* V \\
                             B U - A^* V\end{array}\right).
    \end{array}
    \end{eqnarray}

    This compares perfectly well with the standard eigenstates of the static solution $\mathcal{R}_0$ (\ref{ATDHFB_wf}), which appear in pairs with
    eigenvalues 0 and 1,
    \begin{eqnarray}\label{ATDHFB_final2}
        \mathcal{R}_0 
        &=&
        \bigg(\begin{array}{cr}
            A &  B^*\\
            B &  A^*\\
        \end{array}\bigg)
        \bigg(\begin{array}{cc}
            0      &  0     \\
            0      &  1      \\
        \end{array}\bigg)
        \bigg(\begin{array}{cr}
            A^+ &  B^+\\
            B^T &  A^T\\
        \end{array}\bigg)
        \nonumber \\
        &=&
        \bigg(\begin{array}{cc}
            \rho_0 &  \kappa_0\\
       -\kappa_0^* & 1-\rho_0^*\\
        \end{array}\bigg) =
        \varphi\varphi^+ .
    \end{eqnarray}
    We see that the zero-order density matrix $\mathcal{R}_0$~(\ref{ATDHFB_final2}) is given as a sum of the zero-order quasihole wave functions $\varphi$~(\ref{ATDHFB_wf}) with all quasioccupation factors equal to 1. Similarly, the first-order density matrix $\tilde{\mathcal{R}}_1$~(\ref{ATDHFB_final}) is given as a sum of both the first-order quasiparticle and quasihole wave functions $\chi_1$ and $\varphi_1$~(\ref{ATDHFB_wf1})--(\ref{ATDHFB_wf2}) with the opposite quasioccupation factors of $+\omega_\mu$ and  $-\omega_\mu$, respectively.
    
This enables an easy adaptation of the standard iteration of the HFB equation~\cite{Ring2004book} to the novel iterative solution of the ATDHFB equation presented in Eq.~(\ref{ATDHFB_EQ_it_qp}). Indeed, we first determine the time-odd density matrix $\rho_1$ and the time-odd pairing tensor $\kappa_1$ in Eq.~(\ref{ATDHFB_final}) based on the quasiparticle wave functions (\ref{ATDHFB_wf1}) and  (\ref{ATDHFB_wf2}). Second, we employ the standard HFB algorithm~\cite{Ring2004book} to determine the time-odd mean field $\Gamma_1$ and the time-odd pairing field $\Delta_1$ from $\rho_1$ and $\kappa_1$, leading to $\tilde{\mathcal{H}}_1=\left(\begin{array}{cc}
    \Gamma_1      &  \Delta_1     \\
    -\Delta_1^*      &  -\Gamma_1^*  \\
    \end{array}\right)$ and $E_1 = \varphi^{+}\tilde{\mathcal{H}}_1\chi$. Third, we utilise Eq.~(\ref{ATDHFB_EQ_it_qp}) to determine the matrix $Z$ for the next iteration. And fourth, the SVD of the matrix $Z$ enables the closure of the self-consistent loop by providing the quasiparticle wave functions (\ref{ATDHFB_wf1}) and (\ref{ATDHFB_wf2}).
        
    We see that the iterative equations (\ref{ATDHFB_EQ_it}) and (\ref{ATDHFB_EQ_it_qp}) involve only one-body quasiparticle matrices, and the stability matrix does not need to be evaluated or inverted. Obviously, in the first iteration, in which $\tilde{\mathcal{H}}_1$ is neglected, $\mathcal{M}^{(1)}$ gives the IB collective inertia.
    
    The method was implemented in the non-relativistic density functional solver HFODD~\cite{(Dob21f),(Dob25)}, which solves the nuclear DFT equations by expanding the quasiparticle wavefunction on a three-dimensional Cartesian harmonic-oscillator basis and allows for the self-consistent description of nuclei with arbitrary shapes. The precision of the iterative method was verified by comparing the calculated Thouless-Valatin moments of inertia with those obtained from cranking calculations, yielding perfect agreement~\cite {(Sun25a),Sun2025}.
    
    \begin{figure}[t]
        \centering
        \includegraphics[width=1.0\linewidth]{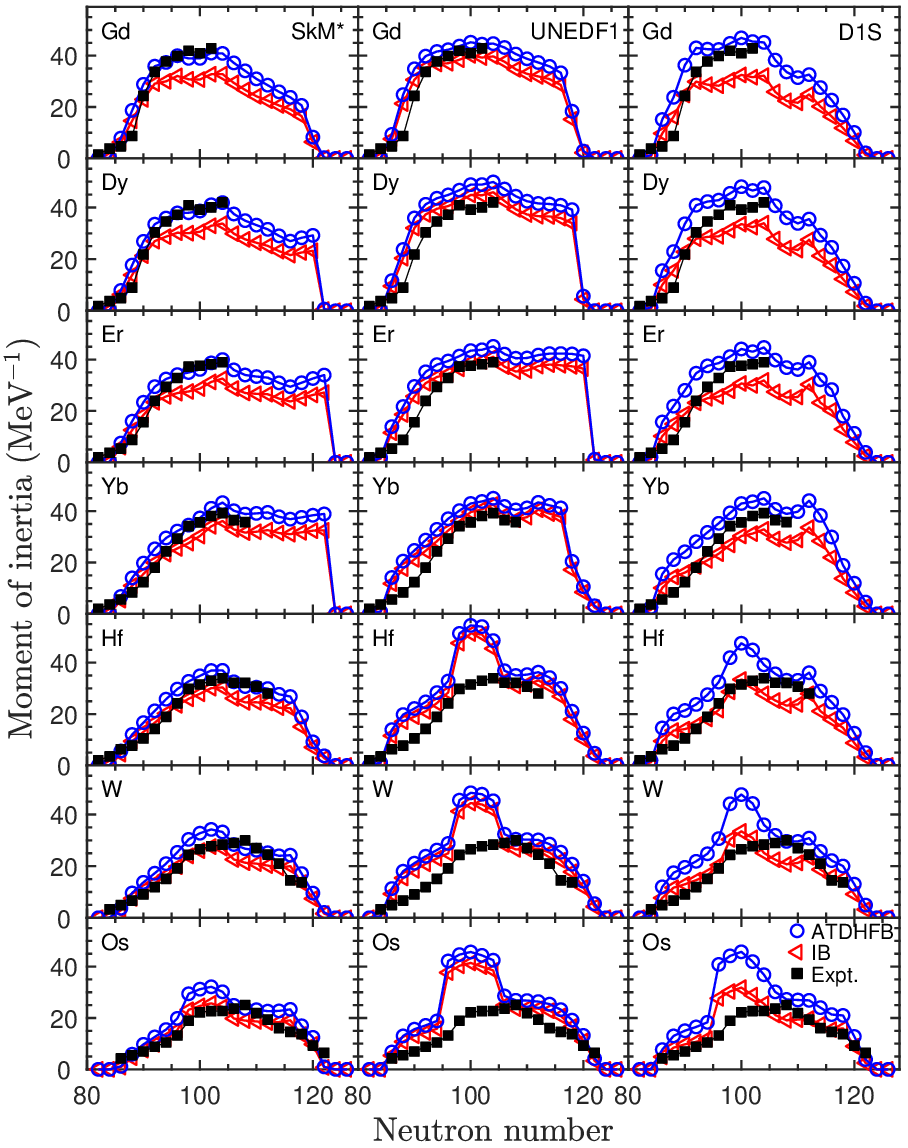}
        \caption{Calculated ATDHFB (circles) and IB (triangles) moments of inertia compared with the values extracted from the experimental data (squares)~\cite{nndc}.}
        \label{fig_gs2os}
    \end{figure}    
    
    The IB collective inertia has been widely used in multiple studies of nuclear collective motion. However, it was soon realised that the calculated values were too low to describe the data. As a result, in most applications, the IB values were multiplied by ad hoc factors of about 1.3, see, e.g., recent Refs.~\cite{Giuliani2018,VRETENAR2005101,Ryssens2022} and references cited therein. 
    Those were purely empirical factors meant to compensate for the unknown effects of the time-odd mean fields. In the present work, having a rapid and efficient method to include those effects, we microscopically determined the ratios of the ATDHFB and IB moments of inertia and compared the full ATDHFB results with the data.
    
    To this end, we performed systematic ATDHFB calculations for the Gd, Dy, Er, Yb, Hf, W, and Os isotopes with neutron numbers between 82 and 126. Two different Skyrme functionals, SkM*~\cite{BARTEL198279} and UNEDF1~\cite{Kortelainen2012PRC},  and one Gogny functional D1S~\cite{BERGER198985} were used for this purpose. For SkM*, the volume pairing with the neutron and proton strengths of $-$178.83 and $-$211.20\,MeV\,fm$^3$~\cite{Petrik2018PRC} were used, respectively. For UNEDF1, the mixed pairing of the type $V_0^t(1-\rho/\rho_{\textrm{sat}})$ was used with the neutron and proton parameters $V_0^n = -223.278$ and $V_0^p = -247.896$\,MeV\,fm$^3$, and with $\rho_{\textrm{sat}}=0.32~\textrm{fm}^{-3}$. Since the time-odd sectors of those Skyrme functionals were originally not adjusted to data, here we fixed them by employing the adjustments performed in Ref.~\cite{(Sas22c)}.
    
    In Fig.~\ref{fig_gs2os}, the calculated ATDHFB and IB moments of inertia are compared with the values extracted from the experimental data using the measured values of the first ${2^+}$ excitation energies \cite{nndc} and the rotational model formula,
    %\begin{equation}
    $\mathcal{I}^{\textrm{Expt.}} = 3\hbar^2/E^1_{2^+}$.
    %\end{equation}
    
    Within the set of studied isotopes, the moments of inertia increase with the neutron number and reach maxima around $N=100$ where the neutron $2f_{7/2}$ orbital becomes occupied, and then decrease. In all isotopes, the calculated ATDHFB moments of inertia are larger than the IB moment of inertia as the time-odd mean fields make significant contributions.
    When compared to the SkM* functional, the calculated moments of inertia obtained for UNEDF1 are systematically larger, even though these two Skyrme functionals predict quite similar quadrupole deformations. For example, the IB moment of inertia of $^{166}\textrm{Er}$ are 27.584\,$\hbar^2$/MeV for SkM* and 36.401\,$\hbar^2$/MeV for UNEDF1, the ATDHFB moments of inertia are 34.950 and 41.342\,$\hbar^2$/MeV, whereas the quadrupole deformations $\beta_{20}$ are 0.320 and 0.324, respectively.
    
    Both SkM* and UNEDF1 are very successful Skyrme functionals in predicting nuclear ground state properties. However, as most of the nuclear density functionals are, SkM*, UNEDF1, and D1S are calibrated to experimental masses and charge radii and their parameters relating to nuclear dynamical properties are not fully constrained. Therefore, the predictions of the collective inertia by different functionals vary.
    The Gogny functional D1S again generates a similar quadrupole deformation of $\beta_{20} = 0.322$ for $^{166}\textrm{Er}$, whereas the predicted IB and ATDHFB moments of inertia are 27.865 and 41.147\,$\hbar^2$/MeV, respectively. The large deviations between the IB and ATDHFB moments of inertia indicate that the Gogny functional D1S predicts strong time-odd mean fields for nuclear collective rotation.
    
    \begin{figure}[t]
        \centering
        \includegraphics[width=1.0\linewidth]{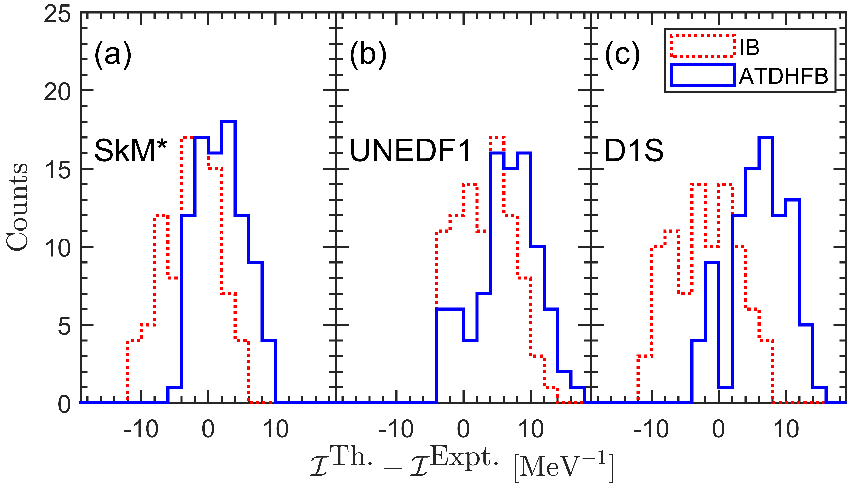}
        \caption{Distributions of the residuals between the IB moments of inertia and the experimental values (dotted line) and between the ATDHFB moments of inertia and the experimental values (solid line) for SkM* (a), UNEDF1 (b), and D1S (c) functionals, respectively.}
        \label{fig_stat}
    \end{figure}
    In Fig.~\ref{fig_stat}, the distribution of the residuals between the calculated moments of inertia and the experimental values $(\mathcal{I}^{\textrm{Th.}}-\mathcal{I}^{\textrm{Expt.}})$ of all the experimentally available isotopes between $N$ = 86 and 122 are presented for SkM*, UNEDF1, and D1S functionals, respectively. 
    The shift between the distribution of the residuals for IB and ATDHFB moments of inertia can be clearly observed, which shows the importance of the time-odd mean field.
    
    Sudden increase of the deformation in the Hf, W and Os isotopes around $N=100$, predicted by UNEDF1 and D1S, is not visible in the experimental data, possibly because triaxiality is neglected in our calculations. Therefore, in the following we present the statistical analysis corresponding to both excluding and including these nuclei, with the latter results given in parentheses.
    
    For SkM*, the mean value of the residual between the IB moment of inertia and the experimental data is $-$2.55 ($-$2.21)\,$\hbar^2$/MeV, that is, for this functional, the IB formula underestimates the moment of inertia. On the contrary, an overestimation is predicted by the ATDHFB method with a mean value of 2.54 (3.06)\,$\hbar^2$/MeV. In both cases, the distributions are fairly narrow with standard deviations of 4.33 (4.21)\,$\hbar^2$/MeV and 3.36 (3.48)\,$\hbar^2$/MeV for IB and ATDHFB, respectively. For D1S, the IB formula also underestimates the moment of inertia while the ATDHFB method overestimates them, but the distributions of the residual are wider than for SkM*. For UNEDF1, wider distributions are also found but both the IB formula and ATDHFB method overestimate the experimental values.
    
    \begin{figure}[t]
        \centering
        \includegraphics[width=0.9\linewidth]{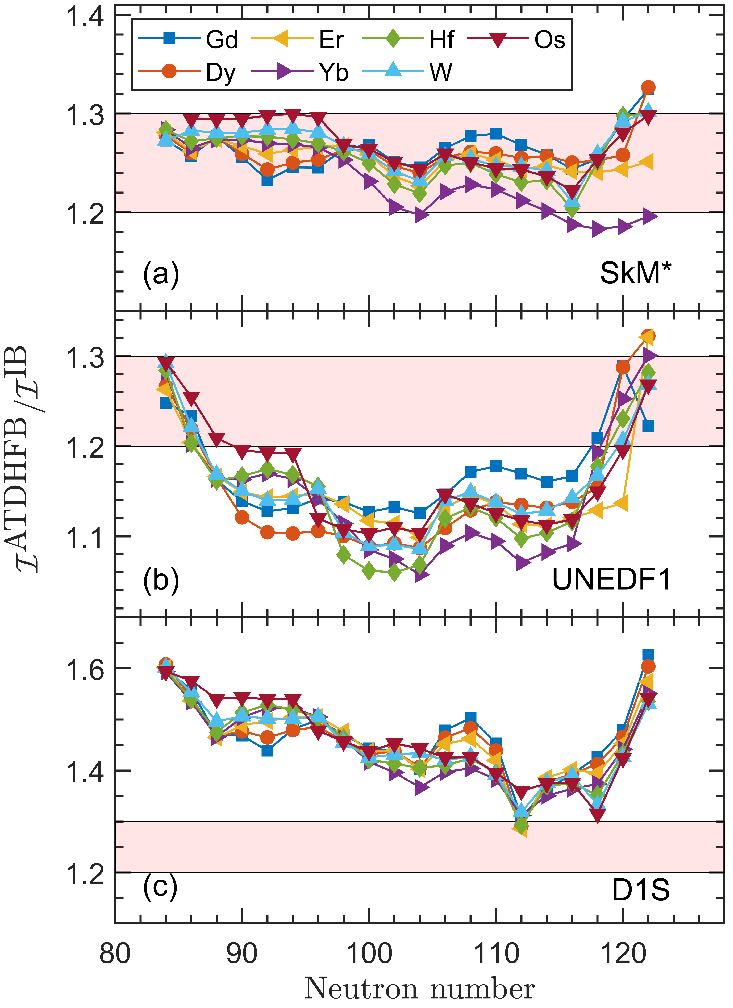}
        \caption{Ratios of the ATDHFB and IB moments of inertia obtained for the SkM* (a), UNEDF1 (b) and D1S (c) functionals compared with the empirical enhancement factors of 1.2--1.3 (shaded areas).}
        \label{ratioSKM}
    \end{figure}
    It should be stressed that for the Skyrme functionals SkM* and UNEDF1, there is some uncertainty in the form and strength of the pairing force, which may consistently shift the IB and ATDHFB moments of inertia up or down. Therefore, firm conclusions on whether the IB or ATDHFB values describe the data better are not possible. Even though an analogous uncertainty does not apply to D1S, as its pairing properties are fixed by the parameters of the functional, the results are inconclusive here too. Therefore, we now proceed to directly compare the IB and ATDHFB moments of inertia to delineate the role of the time-odd mean fields in the adiabatic collective rotational motion.
    
    In Fig.~\ref{ratioSKM}, we display the ratios of the ATDHFB and IB moments of inertia calculated with the Skyrme functionals SkM* and UNEDF1, and Gogny functional D1S. For SkM*, the ratios mostly lie consistently between 1.2 and 1.3, with a few exceptions in the even isotopes of $^{186-192}\textrm{Yb}$, $^{186}\textrm{Gd}$, and $^{188}\textrm{Dy}$. For UNEDF1, however, the situation is entirely different. The ratios are in the region of 1.2--1.3 only if the neutron number is very close to the magic numbers 82 and 126, whereas in the mid-shell region, the ratios become significantly smaller. For D1S, large deviations from the region of 1.2--1.3 are found, that is, for the osmium isotopes with neutron numbers from 84 to 94, the ratios could be as large as 1.5 or 1.6. Moreover, for all studied functionals, the ratios heavily depend on the elements and also vary with neutron numbers. Therefore, no single value of the overall enhancement factor between ATDHFB and IB moment of inertia could be justified.
    
\begin{figure}[t]
        \centering
        \includegraphics[width=0.9\linewidth]{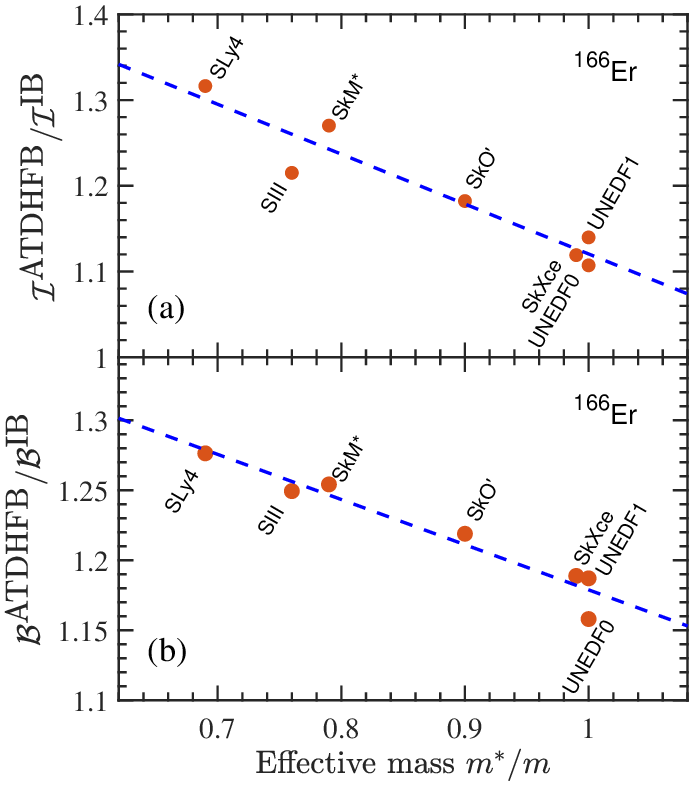}
        \caption{Correlation between the effective mass and the ratio between the ATDHFB and the IB rotational (a) and vibrational (b) collective inertia in $^{166}$Er, calculated with various Skyrme functionals.}
        \label{ratiomg}
    \end{figure}

    To quantitatively clarify the dependence of results on density functionals, we used the seven commonly used Skyrme density functionals (SIII~\cite{Beiner1975}, SkM*~\cite{BARTEL198279}, SkXce~\cite{AlexBrown1998}, SkO'~\cite{Reinhard1999}, SLy4~\cite{Chabanat1998},  UNEDF0~\cite{Kortelainen2010}, and UNEDF1~\cite{Kortelainen2012PRC}) with volume pairing to scrutinize the ratios ATDHFB/IB of rotational and axial vibrational collective inertia in $^{166}$Er, see Fig.~\ref{ratiomg}. The rotational and vibrational collective inertia ratios are linearly correlated with the isoscalar effective masses $m^*$ that characterize given functionals and decrease with slopes of $-$0.582 and $-$0.323, respectively. The correlations are quite strong with coefficients of determination ($R^2$) of 0.9027 and 0.9352. This shows that for these two collective modes, the impact of the time-odd fields on nuclear collective inertia proceeds through the isoscalar current terms $\bm{j}^2$ of the functionals, $\rho\tau-\bm{j}^2$~\cite{Beiner1975}, which are linked to the effective-mass terms $\rho\tau$ by the gauge or Galilean invariance constraints. Nevertheless, even at $m^*/m$=1, the ratios ATDHFB/IB are larger than one, which shows that other time-odd mean fields also play a non-negligible role.

    In summary,	we proposed a novel iterative method to solve the ATDHFB equations exactly and we applied it to investigate the rotational moments of inertia in deformed nuclei. 
    For two Skyrme functionals SkM* and UNEDF1 and one Gogny functional D1S, we determined the rotational moments of inertia of the even-even nuclei between gadolinium and osmium. We obtained a good overall agreement with experimental data, although improvements in functional parameterizations, especially in the pairing channel, are required. The efficiency of our method bids well for the possibility of including the rotational moments of inertia in future adjustments of nuclear functionals.
    
    The focus of our study was on comparing the ATDHFB and Inglis-Belyaev (IB) moments of inertia and gauging the impact of the time-odd mean fields on the nuclear collective motion. We showed that using a fixed numerical multiplicative factor to model the ATDHFB inertia in terms of the IB approximation is not justified. Our results show that the ratios ATDHFB/IB of moments of inertia significantly depend on neutron and proton numbers, deformations, and functionals. 
    
    \section*{Data availability}
    Data will be made available on request.
    
    \section*{Acknowledgements}
    This work was partially supported by the STFC Grant Nos.~ST/P003885/1 and ~ST/V001035/1 and by the Polish National Science Centre under Contract No.~2018/31/ B/ST2/02220. We acknowledge the CSC-IT Center for Science Ltd., Finland, for the allocation of computational resources. This project was partly undertaken on the Viking Cluster, which is a high-performance computing facility provided by the University of York. We are grateful for computational support from the University of York High-Performance Computing service, Viking, and the Research Computing team. We thank Grammarly\textsuperscript{\textregistered} for its support with English writing.

    %\section*{References}	
    \bibliographystyle{elsarticle-num}
    \bibliography{ADB_PLB,jacwit42}

\begin{thebibliography}{10}
\expandafter\ifx\csname url\endcsname\relax
  \def\url#1{\texttt{#1}}\fi
\expandafter\ifx\csname urlprefix\endcsname\relax\def\urlprefix{URL }\fi
\expandafter\ifx\csname href\endcsname\relax
  \def\href#1#2{#2} \def\path#1{#1}\fi

\bibitem{BARANGER1978123}
M.~Baranger, M.~Vénéroni,
  \href{https://www.sciencedirect.com/science/article/pii/0003491678902658}{An
  adiabatic time-dependent {Hartree-Fock} theory of collective motion in finite
  systems}, Annals of Physics 114~(1) (1978) 123--200.
\newblock \href
  {http://dx.doi.org/https://doi.org/10.1016/0003-4916(78)90265-8}
  {\path{doi:https://doi.org/10.1016/0003-4916(78)90265-8}}.
\newline\urlprefix\url{https://www.sciencedirect.com/science/article/pii/0003491678902658}

\bibitem{Ring2004book}
P.~Ring, P.~Schuck, \href{https://link.springer.com/book/9783540212065}{The
  Nuclear Many-Body Problem}, Springer, 2004.
\newline\urlprefix\url{https://link.springer.com/book/9783540212065}

\bibitem{Baran2011}
A.~Baran, J.~A. Sheikh, J.~Dobaczewski, W.~Nazarewicz, A.~Staszczak,
  \href{https://link.aps.org/doi/10.1103/PhysRevC.84.054321}{Quadrupole
  collective inertia in nuclear fission: Cranking approximation}, Phys. Rev. C
  84 (2011) 054321.
\newblock \href {http://dx.doi.org/10.1103/PhysRevC.84.054321}
  {\path{doi:10.1103/PhysRevC.84.054321}}.
\newline\urlprefix\url{https://link.aps.org/doi/10.1103/PhysRevC.84.054321}

\bibitem{Inglis1956PRC}
D.~R. Inglis, Nuclear moments of inertia due to nucleon motion in a rotating
  well, Phys. Rev. 103 (1956) 1786--1795.
\newblock \href {http://dx.doi.org/10.1103/PhysRev.103.1786}
  {\path{doi:10.1103/PhysRev.103.1786}}.

\bibitem{BELIAEV1961NP}
S.~Beliaev,
  \href{https://www.sciencedirect.com/science/article/pii/0029558261903844}{Concerning
  the calculation of the nuclear moment of inertia}, Nuclear Physics 24~(2)
  (1961) 322--325.
\newblock \href
  {http://dx.doi.org/https://doi.org/10.1016/0029-5582(61)90384-4}
  {\path{doi:https://doi.org/10.1016/0029-5582(61)90384-4}}.
\newline\urlprefix\url{https://www.sciencedirect.com/science/article/pii/0029558261903844}

\bibitem{Goutte2005}
H.~Goutte, J.~F. Berger, P.~Casoli, D.~Gogny,
  \href{https://link.aps.org/doi/10.1103/PhysRevC.71.024316}{Microscopic
  approach of fission dynamics applied to fragment kinetic energy and mass
  distributions in $^{238}\mathrm{U}$}, Phys. Rev. C 71 (2005) 024316.
\newblock \href {http://dx.doi.org/10.1103/PhysRevC.71.024316}
  {\path{doi:10.1103/PhysRevC.71.024316}}.
\newline\urlprefix\url{https://link.aps.org/doi/10.1103/PhysRevC.71.024316}

\bibitem{Wen2022}
K.~Wen, T.~Nakatsukasa,
  \href{https://link.aps.org/doi/10.1103/PhysRevC.105.034603}{Microscopic
  collective inertial masses for nuclear reaction in the presence of nucleonic
  effective mass}, Phys. Rev. C 105 (2022) 034603.
\newblock \href {http://dx.doi.org/10.1103/PhysRevC.105.034603}
  {\path{doi:10.1103/PhysRevC.105.034603}}.
\newline\urlprefix\url{https://link.aps.org/doi/10.1103/PhysRevC.105.034603}

\bibitem{Nik2009PRC}
T.~Nik\ifmmode \check{s}\else \v{s}\fi{}i\ifmmode~\acute{c}\else \'{c}\fi{},
  Z.~P. Li, D.~Vretenar, L.~Pr\'ochniak, J.~Meng, P.~Ring,
  \href{https://link.aps.org/doi/10.1103/PhysRevC.79.034303}{Beyond the
  relativistic mean-field approximation. {III. Collective Hamiltonian} in five
  dimensions}, Phys. Rev. C 79 (2009) 034303.
\newblock \href {http://dx.doi.org/10.1103/PhysRevC.79.034303}
  {\path{doi:10.1103/PhysRevC.79.034303}}.
\newline\urlprefix\url{https://link.aps.org/doi/10.1103/PhysRevC.79.034303}

\bibitem{Sadhukhan2013}
J.~Sadhukhan, K.~Mazurek, A.~Baran, J.~Dobaczewski, W.~Nazarewicz, J.~A.
  Sheikh,
  \href{https://link.aps.org/doi/10.1103/PhysRevC.88.064314}{Spontaneous
  fission lifetimes from the minimization of self-consistent collective
  action}, Phys. Rev. C 88 (2013) 064314.
\newblock \href {http://dx.doi.org/10.1103/PhysRevC.88.064314}
  {\path{doi:10.1103/PhysRevC.88.064314}}.
\newline\urlprefix\url{https://link.aps.org/doi/10.1103/PhysRevC.88.064314}

\bibitem{Libert1999PRC}
J.~Libert, M.~Girod, J.-P. Delaroche,
  \href{https://link.aps.org/doi/10.1103/PhysRevC.60.054301}{Microscopic
  descriptions of superdeformed bands with the {Gogny} force: {Configuration}
  mixing calculations in the {A}$\ensuremath{\sim}$190 mass region}, Phys. Rev.
  C 60 (1999) 054301.
\newblock \href {http://dx.doi.org/10.1103/PhysRevC.60.054301}
  {\path{doi:10.1103/PhysRevC.60.054301}}.
\newline\urlprefix\url{https://link.aps.org/doi/10.1103/PhysRevC.60.054301}

\bibitem{ZPLI2012PRC}
Z.~P. Li, T.~Nik\ifmmode \check{s}\else \v{s}\fi{}i\ifmmode~\acute{c}\else
  \'{c}\fi{}, P.~Ring, D.~Vretenar, J.~M. Yao, J.~Meng,
  \href{https://link.aps.org/doi/10.1103/PhysRevC.86.034334}{Efficient method
  for computing the {Thouless-Valatin} inertia parameters}, Phys. Rev. C 86
  (2012) 034334.
\newblock \href {http://dx.doi.org/10.1103/PhysRevC.86.034334}
  {\path{doi:10.1103/PhysRevC.86.034334}}.
\newline\urlprefix\url{https://link.aps.org/doi/10.1103/PhysRevC.86.034334}

\bibitem{Petrik2018PRC}
K.~Petr\'{\i}k, M.~Kortelainen,
  \href{https://link.aps.org/doi/10.1103/PhysRevC.97.034321}{{Thouless-Valatin}
  rotational moment of inertia from linear response theory}, Phys. Rev. C 97
  (2018) 034321.
\newblock \href {http://dx.doi.org/10.1103/PhysRevC.97.034321}
  {\path{doi:10.1103/PhysRevC.97.034321}}.
\newline\urlprefix\url{https://link.aps.org/doi/10.1103/PhysRevC.97.034321}

\bibitem{(Was24)}
K.~Washiyama, N.~Hinohara, T.~Nakatsukasa,
  \href{https://link.aps.org/doi/10.1103/PhysRevC.109.L051301}{Five-dimensional
  collective {Hamiltonian} with improved inertial functions}, Phys. Rev. C 109
  (2024) L051301.
\newblock \href {http://dx.doi.org/10.1103/PhysRevC.109.L051301}
  {\path{doi:10.1103/PhysRevC.109.L051301}}.
\newline\urlprefix\url{https://link.aps.org/doi/10.1103/PhysRevC.109.L051301}

\bibitem{(Sun24)}
X.~Sun, J.~Dobaczewski, M.~Kortelainen, D.~Muir, J.~Sadhukhan,
  A.~Sánchez-Fernández, H.~Wibowo,
  \href{https://arxiv.org/abs/2411.18404}{Iterative solutions of the {ATDHFB}
  equations to determine the nuclear collective inertia} (2024).
\newblock \href {http://arxiv.org/abs/2411.18404} {\path{arXiv:2411.18404}}.
\newline\urlprefix\url{https://arxiv.org/abs/2411.18404}

\bibitem{Nakatsukasa2016RMP}
T.~Nakatsukasa, K.~Matsuyanagi, M.~Matsuo, K.~Yabana,
  \href{https://link.aps.org/doi/10.1103/RevModPhys.88.045004}{Time-dependent
  density-functional description of nuclear dynamics}, Rev. Mod. Phys. 88
  (2016) 045004.
\newblock \href {http://dx.doi.org/10.1103/RevModPhys.88.045004}
  {\path{doi:10.1103/RevModPhys.88.045004}}.
\newline\urlprefix\url{https://link.aps.org/doi/10.1103/RevModPhys.88.045004}

\bibitem{Skyrme1958}
T.~Skyrme,
  \href{https://www.sciencedirect.com/science/article/pii/0029558258903456}{The
  effective nuclear potential}, Nuclear Physics 9~(4) (1958) 615--634.
\newblock \href
  {http://dx.doi.org/https://doi.org/10.1016/0029-5582(58)90345-6}
  {\path{doi:https://doi.org/10.1016/0029-5582(58)90345-6}}.
\newline\urlprefix\url{https://www.sciencedirect.com/science/article/pii/0029558258903456}

\bibitem{Robledo2018}
L.~M. Robledo, T.~R. Rodríguez, R.~R. Rodríguez-Guzmán,
  \href{https://dx.doi.org/10.1088/1361-6471/aadebd}{Mean field and beyond
  description of nuclear structure with the {Gogny} force: a review}, Journal
  of Physics G: Nuclear and Particle Physics 46~(1) (2018) 013001.
\newblock \href {http://dx.doi.org/10.1088/1361-6471/aadebd}
  {\path{doi:10.1088/1361-6471/aadebd}}.
\newline\urlprefix\url{https://dx.doi.org/10.1088/1361-6471/aadebd}

\bibitem{Sun2025}
{Xuwei Sun {\it et al.}}, to be published.

\bibitem{(Dob81a)}
J.~Dobaczewski, J.~Skalski,
  \href{http://www.sciencedirect.com/science/article/pii/0375947481900105}{The
  quadrupole vibrational inertial function in the adiabatic time-dependent
  {Hartree-Fock-Bogolyubov} approximation}, Nucl. Phys. A 369~(1) (1981) 123 --
  140.
\newblock \href {http://dx.doi.org/10.1016/0375-9474(81)90010-5}
  {\path{doi:10.1016/0375-9474(81)90010-5}}.
\newline\urlprefix\url{http://www.sciencedirect.com/science/article/pii/0375947481900105}

\bibitem{(Dob21f)}
J.~Dobaczewski, P.~B\c{a}czyk, P.~Becker, M.~Bender, K.~Bennaceur, J.~Bonnard,
  Y.~Gao, A.~Idini, M.~Konieczka, M.~Kortelainen, L.~Pr{\'o}chniak, A.~M.
  Romero, W.~Satu{\l}a, Y.~Shi, L.~F. Yu, T.~R. Werner,
  \href{https://doi.org/10.1088/1361-6471/ac0a82}{Solution of universal
  nonrelativistic nuclear {DFT} equations in the {Cartesian} deformed
  harmonic-oscillator basis. {(IX) HFODD} (v3.06h): a new version of the
  program}, J. Phys. G: Nucl. Part. Phys. 48~(10) (2021) 102001.
\newblock \href {http://dx.doi.org/10.1088/1361-6471/ac0a82}
  {\path{doi:10.1088/1361-6471/ac0a82}}.
\newline\urlprefix\url{https://doi.org/10.1088/1361-6471/ac0a82}

\bibitem{(Dob25)}
{J. Dobaczewski {\it et al.}}, Code {\sc hfodd}, version to be published
  (2025).

\bibitem{Giuliani2018}
S.~A. Giuliani, L.~M. Robledo,
  \href{https://www.sciencedirect.com/science/article/pii/S0370269318308165}{Non-perturbative
  collective inertias for fission: A comparative study}, Physics Letters B 787
  (2018) 134--140.
\newblock \href
  {http://dx.doi.org/https://doi.org/10.1016/j.physletb.2018.10.045}
  {\path{doi:https://doi.org/10.1016/j.physletb.2018.10.045}}.
\newline\urlprefix\url{https://www.sciencedirect.com/science/article/pii/S0370269318308165}

\bibitem{VRETENAR2005101}
D.~Vretenar, A.~Afanasjev, G.~Lalazissis, P.~Ring,
  \href{https://www.sciencedirect.com/science/article/pii/S0370157304004545}{Relativistic
  {Hartree-Bogoliubov} theory: static and dynamic aspects of exotic nuclear
  structure}, Physics Reports 409~(3) (2005) 101--259.
\newblock \href
  {http://dx.doi.org/https://doi.org/10.1016/j.physrep.2004.10.001}
  {\path{doi:https://doi.org/10.1016/j.physrep.2004.10.001}}.
\newline\urlprefix\url{https://www.sciencedirect.com/science/article/pii/S0370157304004545}

\bibitem{Ryssens2022}
W.~Ryssens, G.~Scamps, S.~Goriely, M.~Bender,
  \href{https://doi.org/10.1140/epja/s10050-022-00894-5}{{Skyrme-Hartree-Fock-Bogoliubov}
  mass models on a {3D mesh: II. Time-reversal} symmetry breaking}, The
  European Physical Journal A 58~(12) (2022) 246.
\newblock \href {http://dx.doi.org/10.1140/epja/s10050-022-00894-5}
  {\path{doi:10.1140/epja/s10050-022-00894-5}}.
\newline\urlprefix\url{https://doi.org/10.1140/epja/s10050-022-00894-5}

\bibitem{BARTEL198279}
J.~Bartel, P.~Quentin, M.~Brack, C.~Guet, H.-B. Håkansson,
  \href{https://www.sciencedirect.com/science/article/pii/0375947482904031}{Towards
  a better parametrisation of {Skyrme-like} effective forces: {A critical study
  of the SkM} force}, Nuclear Physics A 386~(1) (1982) 79--100.
\newblock \href
  {http://dx.doi.org/https://doi.org/10.1016/0375-9474(82)90403-1}
  {\path{doi:https://doi.org/10.1016/0375-9474(82)90403-1}}.
\newline\urlprefix\url{https://www.sciencedirect.com/science/article/pii/0375947482904031}

\bibitem{Kortelainen2012PRC}
M.~Kortelainen, J.~McDonnell, W.~Nazarewicz, P.-G. Reinhard, J.~Sarich,
  N.~Schunck, M.~V. Stoitsov, S.~M. Wild,
  \href{https://link.aps.org/doi/10.1103/PhysRevC.85.024304}{Nuclear energy
  density optimization: Large deformations}, Phys. Rev. C 85 (2012) 024304.
\newblock \href {http://dx.doi.org/10.1103/PhysRevC.85.024304}
  {\path{doi:10.1103/PhysRevC.85.024304}}.
\newline\urlprefix\url{https://link.aps.org/doi/10.1103/PhysRevC.85.024304}

\bibitem{BERGER198985}
J.~Berger, M.~Girod, D.~Gogny,
  \href{https://www.sciencedirect.com/science/article/pii/0375947489906568}{Constrained
  {Hartree-Fock} and beyond}, Nuclear Physics A 502 (1989) 85--104.
\newblock \href
  {http://dx.doi.org/https://doi.org/10.1016/0375-9474(89)90656-8}
  {\path{doi:https://doi.org/10.1016/0375-9474(89)90656-8}}.
\newline\urlprefix\url{https://www.sciencedirect.com/science/article/pii/0375947489906568}

\bibitem{(Sas22c)}
P.~L. Sassarini, J.~Dobaczewski, J.~Bonnard, R.~F. {Garcia Ruiz},
  \href{https://doi.org/10.1088/1361-6471/ac900a}{Nuclear {DFT} analysis of
  electromagnetic moments in odd near doubly magic nuclei}, Journal of Physics
  G: Nuclear and Particle Physics 49~(11) (2022) 11LT01.
\newblock \href {http://dx.doi.org/10.1088/1361-6471/ac900a}
  {\path{doi:10.1088/1361-6471/ac900a}}.
\newline\urlprefix\url{https://doi.org/10.1088/1361-6471/ac900a}

\bibitem{nndc}
Brookhaven National Laboratory, \href{https://www.nndc.bnl.gov/nudat3/}{NuDat
  3, National Nuclear Data Center (NNDC)}.
\newline\urlprefix\url{https://www.nndc.bnl.gov/nudat3/}

\bibitem{Beiner1975}
M.~Beiner, H.~Flocard, N.~{Van Giai}, P.~Quentin,
  \href{https://www.sciencedirect.com/science/article/pii/0375947475903383}{Nuclear
  ground-state properties and self-consistent calculations with the {Skyrme}
  interaction: {(I). Spherical} description}, Nuclear Physics A 238~(1) (1975)
  29--69.
\newblock \href
  {http://dx.doi.org/https://doi.org/10.1016/0375-9474(75)90338-3}
  {\path{doi:https://doi.org/10.1016/0375-9474(75)90338-3}}.
\newline\urlprefix\url{https://www.sciencedirect.com/science/article/pii/0375947475903383}

\bibitem{AlexBrown1998}
B.~Alex~Brown, \href{https://link.aps.org/doi/10.1103/PhysRevC.58.220}{New
  {Skyrme} interaction for normal and exotic nuclei}, Phys. Rev. C 58 (1998)
  220--231.
\newblock \href {http://dx.doi.org/10.1103/PhysRevC.58.220}
  {\path{doi:10.1103/PhysRevC.58.220}}.
\newline\urlprefix\url{https://link.aps.org/doi/10.1103/PhysRevC.58.220}

\bibitem{Reinhard1999}
P.-G. Reinhard,
  \href{https://www.sciencedirect.com/science/article/pii/S0375947499000767}{Skyrme
  forces and giant resonances in exotic nuclei}, Nuclear Physics A 649~(1)
  (1999) 305--314, giant Resonances.
\newblock \href
  {http://dx.doi.org/https://doi.org/10.1016/S0375-9474(99)00076-7}
  {\path{doi:https://doi.org/10.1016/S0375-9474(99)00076-7}}.
\newline\urlprefix\url{https://www.sciencedirect.com/science/article/pii/S0375947499000767}

\bibitem{Chabanat1998}
E.~Chabanat, P.~Bonche, P.~Haensel, J.~Meyer, R.~Schaeffer,
  \href{https://www.sciencedirect.com/science/article/pii/S0375947498001808}{A
  {Skyrme} parametrization from subnuclear to neutron star densities {Part II.
  Nuclei} far from stabilities}, Nuclear Physics A 635~(1) (1998) 231--256.
\newblock \href
  {http://dx.doi.org/https://doi.org/10.1016/S0375-9474(98)00180-8}
  {\path{doi:https://doi.org/10.1016/S0375-9474(98)00180-8}}.
\newline\urlprefix\url{https://www.sciencedirect.com/science/article/pii/S0375947498001808}

\bibitem{Kortelainen2010}
M.~Kortelainen, T.~Lesinski, J.~Mor\'e, W.~Nazarewicz, J.~Sarich, N.~Schunck,
  M.~V. Stoitsov, S.~Wild,
  \href{https://link.aps.org/doi/10.1103/PhysRevC.82.024313}{Nuclear energy
  density optimization}, Phys. Rev. C 82 (2010) 024313.
\newblock \href {http://dx.doi.org/10.1103/PhysRevC.82.024313}
  {\path{doi:10.1103/PhysRevC.82.024313}}.
\newline\urlprefix\url{https://link.aps.org/doi/10.1103/PhysRevC.82.024313}

\end{thebibliography}
    
    %------------------------------------------------
\end{document}